\documentclass[a4paper,11pt]{article}
\usepackage{pos}
\usepackage{tikz}
\usepackage{pgfplots}
\usepackage{graphicx}
\usepackage{caption}
\usepackage{subcaption}
\usepackage{float}
\usepackage[firstpageonly=true,angle=0,hpos=18.0cm,vpos=5.8cm,text=MIT-CTP/5721,vanchor=t,hanchor=l,fontsize=12pt, color=black]{draftwatermark}
\newcommand{\todo}[1]{}
\renewcommand{\todo}[1]{{\color{red} TODO: {#1}}}
\newcommand{\csw}{c_{\mathrm{SW}}}
\newcommand{\mps}{m_{\mathrm{PS}}}
\newcommand{\mv}{m_{\mathrm{V}}}

\newcommand{\tr}{\mathrm{Tr\,}}

\title{The singlet scalar state in a chiral ensemble in $SU(2)$ with two fundamental flavours}
\ShortTitle{The mass of the $\sigma$ in $SU(2)$ with two fundamental flavours}
\author*[a]{Laurence Sebastian Bowes}
\author[a]{Vincent Drach}
\author[b]{Patrick Fritzsch}
\author[c]{Sofie~Martins}
\author[a,c]{Antonio Rago}
\author[d,e]{Fernando Romero-L\'opez}

\affiliation[a]{Centre for Mathematical Sciences, University of Plymouth, England, UK}
\affiliation[b]{School of Mathematics, Trinity College Dublin, Ireland}
\affiliation[c]{IMADA and Quantum Theory Center, University of Southern Denmark, Denmark}
\affiliation[d]{Center for Theoretical Physics, Massachusetts Institute of Technology, USA}
\affiliation[e]{Albert Einstein Center, Institute for Theoretical Physics, University of Bern,  Switzerland}
\emailAdd{laurence.bowes@plymouth.ac.uk}
\emailAdd{vincent.drach@plymouth.ac.uk}
\emailAdd{fritzscp@tcd.ie}
\emailAdd{martinss@imada.sdu.dk}
\emailAdd{rago@qtc.sdu.dk}
\emailAdd{fernando.romero-lopez@unibe.ch}

\abstract{Composite Higgs models are a class of Beyond the Standard Model (BSM) models proposed to address the hierarchy and naturalness problems associated with the Standard Model (SM) Higgs. A new QCD-like strongly interacting sector based on $SU(2)$ with two fundamental flavours can be used to build a composite Higgs model which is not yet ruled out by experiment. The role of the singlet scalar resonance will affect Higgs phenomenology at the LHC. In this project our goal is to understand the properties of the singlet scalar state in the new strongly interacting sector in isolation as a first step to understanding the role of this state in composite Higgs models. We present here the first lattice results for the mass of the $\sigma$ in $SU(2)$ with two fundamental flavours using exponential clover Wilson fermions.}

\FullConference{The 41st International Symposium on Lattice Field Theory (LATTICE2024)\\
 28 July - 3 August 2024\\
Liverpool, UK\\}

\begin{document}
\maketitle
\section{Introduction}

Composite Higgs models are a class of Beyond the Standard Model (BSM) theories that give a dynamical origin to the electroweak symmetry breaking in the Standard Model by introducing a new strongly interacting sector, with a view to explaining the hierarchy and naturalness problems. The new sector is charged under the electroweak interaction such that it gives a dynamical origin to the observed Higgs boson. In the new sector in isolation of the SM there is expected to be a rich spectrum analogous to QCD. Understanding how the spectrum of the new sector affects particle physics phenomenology at the LHC is therefore of great interest.
In particular, we wish to gain a good understanding of scattering processes in the composite Higgs sector, and how these would affect scattering processes currently being probed at the LHC.

A viable theory for the new strongly interacting sector is $SU(2)$ with two fundamental flavours. It features an enhanced chiral symmetry breaking pattern of $SU(4) \rightarrow Sp(4)$ due to the gauge group $SU(2)$ having a pseudoreal fundamental representation.

In this project, we study the lightest flavour isosinglet Lorentz scalar state in the strongly interacting sector in isolation using lattice simulations. For convenience we refer to the lightest state in this channel as the $\sigma$. It is expected that the $\sigma$ state will be a resonance of two pseudo-Nambu-Goldstone bosons in the chiral limit of the new sector. We aim to use lattice calculations to predict from first principles the scattering properties of this scalar resonance. In order to do that we need to identify a regime where the $\sigma$ is a resonance.

Defining $m_{\mathrm{V}}$ as the lightest isotriplet vector state and $m_{\mathrm{PS}}$ as the lightest isotriplet pseudoscalar state, we use $\frac{m_{\mathrm{V}}}{m_{\mathrm{{PS}}}}$ as a convenient parameter for the theory that we use to measure how close we are to the chiral limit, and and to compare with other work. In addition, $\frac{m_{\mathrm{V}}}{m_{\mathrm{{PS}}}}$ is a measure of the scale separation which controls the validity of chiral perturbation theory.

$SU(2)$ with two fundamental flavours has been previously studied on the lattice with a tree level improved Wilson clover action. The $\sigma$ state was investigated using a fully-fledged L{\"u}scher scattering calculation, where it was shown to likely be a stable state up to $\frac{m_{\mathrm{V}}}{m_{\mathrm{{PS}}}} < 2.5$~\cite{Drach_2022}.

In the rest of this report we outline the details of the lattice setup including the tuning of $\csw$ and the generation of the HMC ensemble. We then turn to the details of the calculation of the $\sigma$ effective mass and some comments on the results.

\section{Lattice Setup}
For the gauge sector we use the Wilson plaquette action, and for the fermion sector we use exponential clover improved Wilson fermions~\cite{Francis:2019muy}. The action is
\begin{equation} 
  S = \sum_{x}\left[\frac{\beta}{2} \sum_{\mu < \nu} \mathfrak{Re}\ \mathrm{Tr} [1 - P_{\mu, \nu}(x)] + \sum_y \overline{\psi}(x)D(x,y)\psi(y) \right],
\end{equation}

where the diagonal part of the even-odd preconditioned Wilson-Dirac operator is
 \begin{equation}
D_{ee}(x,y) + D_{oo}(x,y) = (4+m_0) \exp\left[\frac{\csw}{4+m_0}\frac{i}{4}\sigma_{\mu\nu}\widehat{F}_{\mu\nu}(x)\right]\delta_{xy}.
 \end{equation}
Here $\psi = (u, d)^T$, $\beta= 4/g_0^{2}$, $m_0$ is the bare fermion mass, $\widehat{F}_{\mu\nu}$ is the lattice gauge field strength, $\sigma_{\mu\nu} = \frac{i}{2}[\gamma_\mu, \gamma_\nu]$ and $\csw$ is the Sheikholeslami-Wohlert parameter. $D$ is diagonal in flavour space.

We perform simulations on a volume with spatial extent $L$ and time extent $T$, where all quantities are in lattice units. We use periodic boundary conditions for the HMC simulations, and all of the simulations are performed in the HiRep suite, first described in \cite{Del_Debbio_2010}. In order to generate the ensemble used in this work, we used multilevel integration HMC with two levels of Hasenbusch acceleration. A significant amount of computing resources had to be used to tune the Hasenbusch parameters.

To perform the exponential clover simulations with $O(a)$ improvement, the parameter $\csw$ must be tuned non-perturbatively for each value of $\beta$. The tuning procedure and results are outlined in Section \ref{sec:csw}.

To set the scale we use the Wilson gauge flow and the $w_0$ parameter, defined by $W(w_0^2) = 1$ where $W(t) = t \frac{d}{dt}\left\{t^2 \langle E(t) \rangle_\mathrm{sym}\right\}$, following the definitions in \cite{w0paper}.
As for mesonic observables, we use isospin---a subgroup of $Sp(4)$---to classify the states. The basic flavour triplet operator we use is
\begin{equation}
    \mathcal{O}^{\mathrm{triplet}}_\Gamma(x) = \overline{\psi}_{\alpha i c}(x) \Gamma_{\alpha\beta} \tau^3_{ij} \psi_{\beta j c}(x) = \overline{u}_{\alpha c}(x)\Gamma_{\alpha\beta}u_{\beta c}(x) - \overline{d}_{\alpha c}(x)\Gamma_{\alpha\beta}d_{\beta c}(x)
\end{equation}

and the flavour singlet operator is
\begin{equation}
\label{eq:singlet_operator}
\mathcal{O}^{\mathrm{singlet}}_\Gamma(x) = \overline{\psi}_{\alpha i c}(x) \Gamma_{\alpha\beta} 1_{ij} \psi_{\beta j c}(x) = \overline{u}_{\alpha c}(x)\Gamma_{\alpha\beta}u_{\beta c}(x) + \overline{d}_{\alpha c}(x)\Gamma_{\alpha\beta}d_{\beta c}(x).
\end{equation}

\section{Simulation Results}
\subsection{Tuning of $\csw$} \label{sec:csw}
\begin{figure}[!h]
\centering
\includegraphics[width=0.73\textwidth]{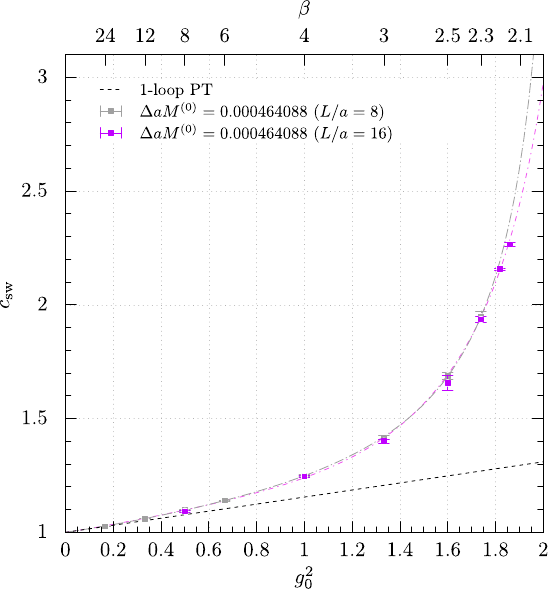}
\caption{The tuned values for $\csw$ against $\beta$. The dashed straight line is the one-loop perturbation theory result.}
\label{fig:csw}
\end{figure}

The value of $\csw$ for each $\beta$ is tuned non-perturbatively via simulations with Schr\"{o}dinger functional boundary conditions following the procedure described in \cite{L_scher_1997}. Figure \ref{fig:csw} shows the final tuned values of $\csw$ against $\beta$ resulting from our extensive simulations. For comparison the one-loop perturbation theory result is also plotted. Simulations at two volumes were performed in order to control the finite volume effects. The results from the larger volume ($L=16$) ensembles are fitted with a Pad\'{e} approximant, where first two terms of the low coupling expansion are matched with the 1-loop perturbation theory result.

A significant amount of effort has been expended on running simulations at $\beta=2.15$, and $\csw$ is now tuned for all values of $\beta \geq 2.15$, which allows us to run at a coarser lattice spacing.  Despite assiduous efforts, simulations at $\beta=2.1$ remain inconclusive. The reader interested in more details about the current status of the tuning of $\csw$ is encouraged to refer to \cite{Sofie}.

\subsection{HMC Ensemble}
\label{sec:ensemble}
\begin{figure}[h]
\includegraphics[width=0.9\textwidth]{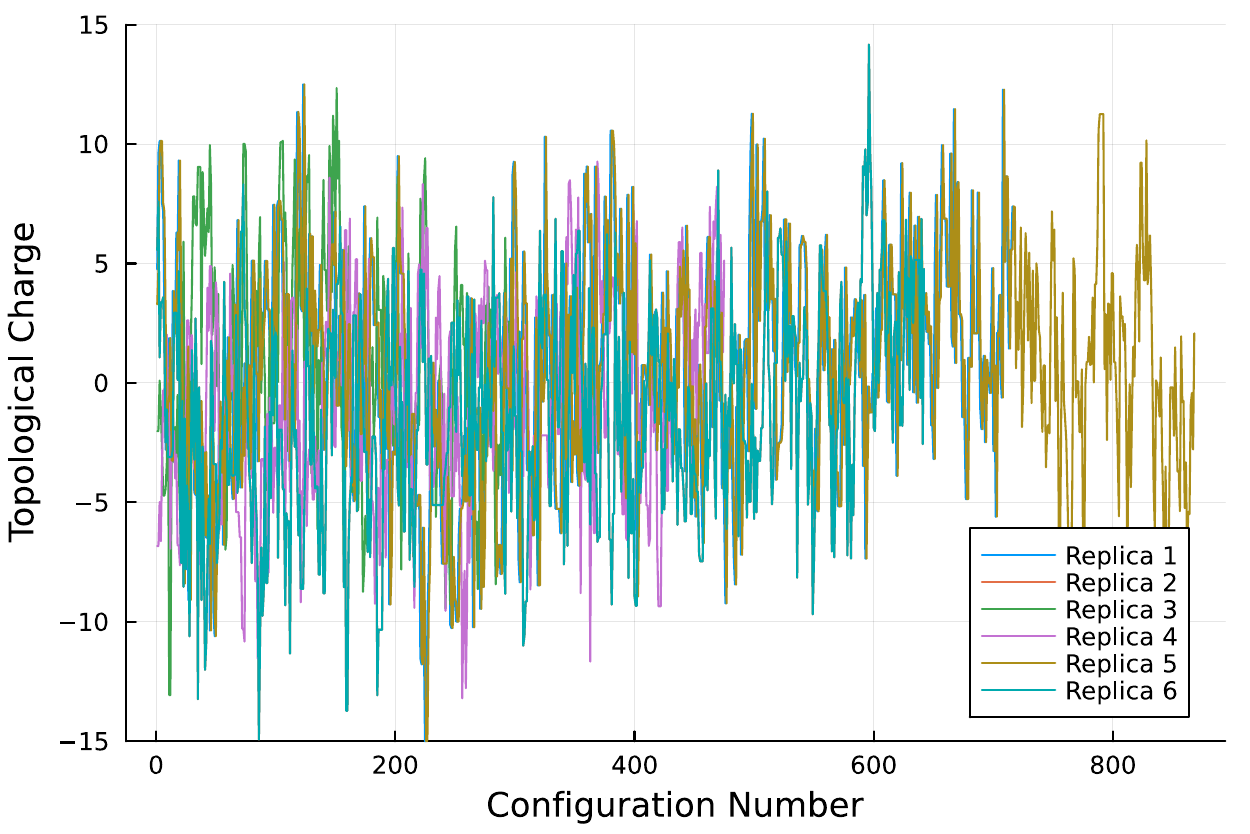}
\caption{Plot of the topological charge history measured at the Wilson flow time $t = w_0^2 = 20.25$.}
\label{fig:tc}
\end{figure}

Based on the results of previously generated heavier mass ensembles at $\beta=2.3$ and $\beta=2.2$ \cite{bowes20232flavoursu2gaugetheory}, we determined by extrapolation that a $\beta=2.2$ ensemble on a volume of $V=64 \times 32^3$ with a bare mass of $m_0 = -0.2864$ would be the lightest we could run at this volume, since it would have $\mps L \approx 5$, which is our chosen cutoff for limiting finite volume effects. The simulation was run on different HPC machines in 6 replicas, and the combined data is shown in Table \ref{tab:ensemble}. The topological charge histories for the replicas are plotted in figure \ref{fig:tc}, and we do not observe any indication of topological freezing.
\begin{table}[!h]
  \begin{center}
\begin{tabular}{||c c c c c c c c c||}
 \hline
 Volume & $\beta$ & $m_0$ & $m_{\mathrm{PS}}$ &$m_{\mathrm{PS}} L $ & $m_{\mathrm{V}}$ & $\frac{m_{\mathrm{V}}}{m_{\mathrm{PS}}}$ & $N_{\mathrm{confs}}$ & $w_0$ \\ [0.5ex]
 \hline
 $64\times 32^3$ & $2.2$ & $-0.2864$ & 0.120(2) & 3.84 & 0.29(2) & 2.46(8)& 3255& 4.50(3)\\
  \hline
\end{tabular}
\end{center}
\caption{Summary data for the ensemble.}
\label{tab:ensemble}
\end{table}

After generating the ensemble, $m_{\mathrm{PS}}$ is measured to be lower than predicted from extrapolation ($\mps L = 3.84$), therefore a more chiral simulation would require a larger volume.

\subsection{Spectroscopy} \label{sec:spec}
\subsubsection{Definitions}
As a first step to understanding the properties of the $\sigma$ resonance on the generated ensemble described in Section \ref{sec:ensemble}, we investigate the behaviour of the two-point function defined as

\begin{equation}
f_\sigma[t] = \langle O_\sigma(t) \overline{O}_\sigma(0) \rangle
\end{equation}
where
\begin{equation}
    O_\sigma(t) = \sum_{\vec{x}}O^{\mathrm{singlet}}_1(x)
\end{equation}
as defined in Equation \ref{eq:singlet_operator}.
Since $O_\sigma(t)$ has the quantum numbers of the vacuum, it has a vacuum expectation value which provides a constant offset to the two-point function. The large $t$ behaviour is
\begin{equation}
f_\sigma[t] \underset{t\,\mathrm{large}}{\longrightarrow} A + B\cosh\left(\left(\frac{T}{2} - t\right)m_\sigma\right)\,.
\end{equation}
where $A$ is the vacuum expectation value, and $m_\sigma$ is the energy of the lightest state in this channel. If the $\sigma$ is a resonance then we expect the fitted effective mass at large $t$ to be $m_\sigma = 2\mps$, and extracting the pole mass of the resonance itself would require a full scattering calculation. To calculate the effective mass of the $\sigma$, we subtract the vacuum expectation value by evaluating it explicitly.

\begin{equation}
    \overline{f}_\sigma(t) = f_\sigma(t) - \langle O_\sigma(t)\rangle \langle O_\sigma(0)\rangle\,.
\end{equation}
The effective mass $m_\mathrm{eff}[t]$ is defined by implicitly solving
\begin{equation} \frac{\overline{f_\sigma}[t-1]}{\overline{f_\sigma}[t]} = \frac{e^{-(T - (t-1))m^\sigma_\mathrm{eff}[t]} + e^{-(t-1)m^\sigma_\mathrm{eff}[t]}}{e^{-(T - t)m^\sigma_\mathrm{eff}[t]} + e^{-tm^\sigma_\mathrm{eff}[t]}}\,.
\end{equation}
The effective mass $m^\sigma_\mathrm{eff}[t]$ is defined such that it plateaus to $m_\sigma$ in the large $t$ limit, and we extract $m_\sigma$ by performing a constant fit for large $t$.

Performing the Wick contractions for the two-point function we obtain
\begin{align}
\begin{split}
   \langle O_\sigma(t) O_\sigma(0) \rangle_F= &\, \frac{1}{\sqrt 2} \sum_{\vec{x}, \vec{y}} \big[ 4 \tr S\{(t, \vec{x}), (t, \vec{x})\} \tr S\{(0, \vec{y}), (0, \vec{y})\}\\
    &- 2 \tr \left[ S\{(0, \vec{y}), (t, \vec{x})\} S\{(t, \vec{x}), (0, \vec{y})\}\right] \big]\,.
    \label{eq:2pf}
\end{split}
\end{align}
$S$ is the gauge dependent fermion propagator $S = D^{-1}$. The first term is a disconnected term, and the second term is a connected term. We evaluate each term separately with different types of stochastic source. In order to evaluate the connected term, we use `spin-explicit wall' (SEMWall) sources. For more details see e.g. \cite{collaboration_2008}. The disconnected term is vastly more expensive to evaluate than the connected term, because it is much more noisy and the next section is dedicated to explaining its evaluation. For both types of stochastic source we use $Z(2) \otimes Z(2)$ noise.
\subsubsection{Evaluation of the disconnected term}

Introduce the `volume' stochastic sources $\xi^k_{\alpha c} (x)$, where $k = 1\ldots N$ and $N$ is the number of stochastic sources, such that for any $x,y,\alpha,\beta, c, d$
\begin{equation}
\lim_{N\rightarrow\infty}\frac{1}{N}\sum_k \xi^{*k}_{\alpha c}(x) \xi^k_{\beta d}(y) = \delta_{\alpha\beta}\delta_{cd}\delta_{xy}.
\end{equation}
If we define
\begin{equation}
\varphi^k_{\alpha c}(x) = \sum_{y, \beta, d} S_{\alpha \beta c d}(x,y)\xi^k_{\beta d}(y)
\end{equation}
where $S$ is the gauge-dependent fermion propagator, it can be shown
\begin{equation}
\tr S\{(t, \vec{x}), (t, \vec{x})\} = \sum_k \xi^{*k}_{\alpha c}(t, \vec{x})\varphi^k_{\alpha c}(t, \vec{x}).
\end{equation}

The disconnected term in Equation \ref{eq:2pf} can then be calculated using two independent sets of stochastic sources $\xi$ and $\tilde{\xi}$ as
\begin{equation}
\label{eq:disceval}
\tr \left[S\{(t, \vec{x}), (t, \vec{x})\}\right] \tr\left[ S\{(0, \vec{y}), (0, \vec{y})\}\right] = \sum_{k,l}\left( \xi^{*k}_{\alpha c}(t, \vec{x})\varphi^k_{\alpha c}(t, \vec{x})\right)\left( \tilde{\xi}^{*l}_{\beta d}(0, \vec{y})\tilde{\varphi}^l_{\beta d}(0, \vec{y})\right).
\end{equation}
We refer to the total number of stochastic inversions as "hits", which are split evenly between each term in Equation \ref{eq:disceval}.

In practice, to improve the signal to noise ratio for the disconnected term, we average over all time separations, defining an improved estimator for the disconnected contribution which we denote

\begin{equation}
\Gamma_\sigma(\Delta t)_\mathrm{disc} = \frac{1}{T} \sum_{t_1-t_2 = \Delta t}(O_\sigma(t_1) O_\sigma(t_2))_\mathrm{disc}\,.
\end{equation}
At this point we add the connected contribution and use the same stochastic sources to construct the gauge-averaged VEV which is then subtracted to provide the full $\sigma$ correlation function. Explicitly,
\begin{equation}
\overline{\Gamma}_\sigma(\Delta t)_\mathrm{disc} = \Gamma_\sigma(\Delta t)_\mathrm{disc} - \left\langle \frac{1}{2NT}\sum_{k=1}^{2N}\sum_{t=0}^{T-1}\sum_{\vec{x},\alpha,c}  \xi^{*k}_{\alpha c}(t, \vec{x})\varphi^k_{\alpha c}(t, \vec{x}) \right\rangle^2.
\end{equation}
\subsubsection{Results}

\begin{figure}[!h]
\centering
\includegraphics[width=0.8\textwidth]{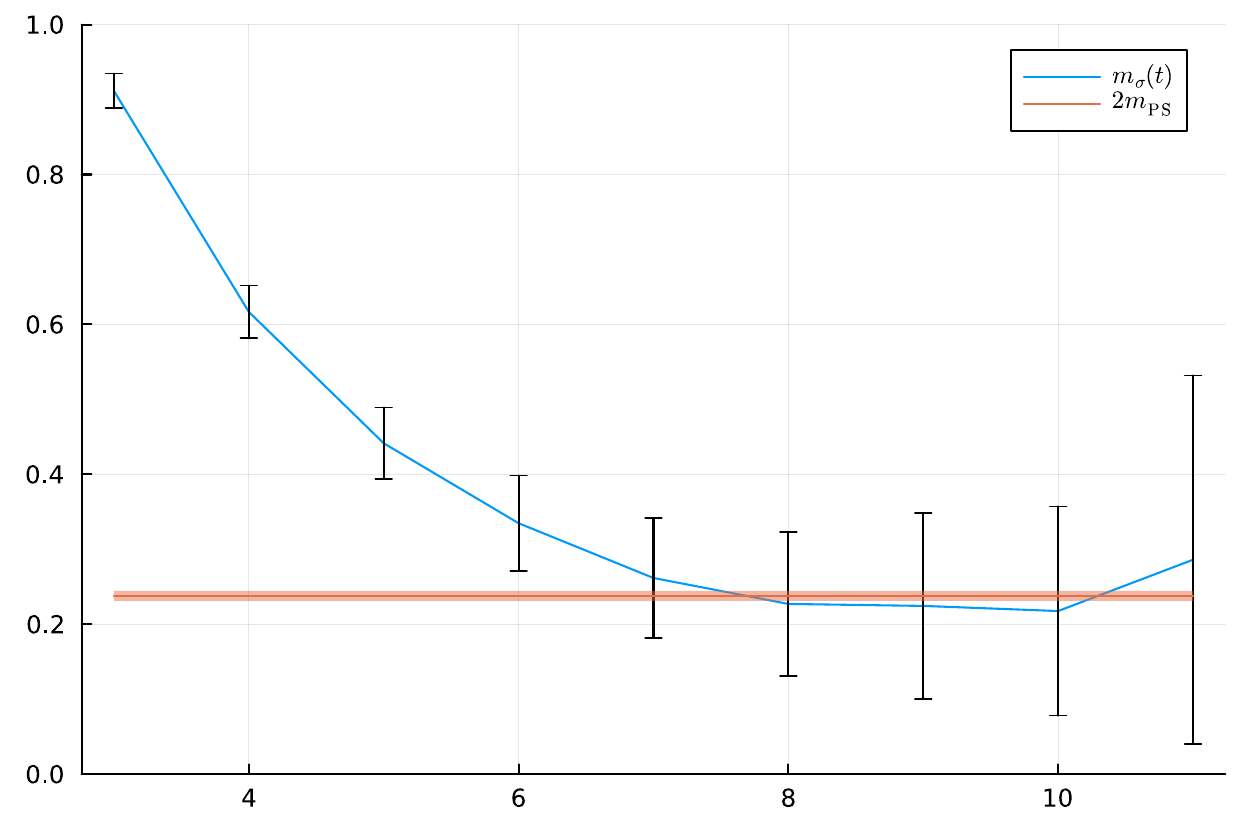}
\caption{The effective mass of the $\sigma$ against lattice time for a binning width of $50$. After timeslice $t = 11$ the effective mass becomes dominated by noise. Also shown is $2m_{\mathrm{PS}}$ in orange. }
\label{fig:effmass}
\end{figure}

The disconnected and connected parts are combined into one correlator, and the effective mass is plotted in Figure \ref{fig:effmass}. The $\sigma$ mass appears compatible with $2\mps$ in this ensemble.

\begin{figure}[!h]
\centering
\includegraphics[width=0.75\textwidth]{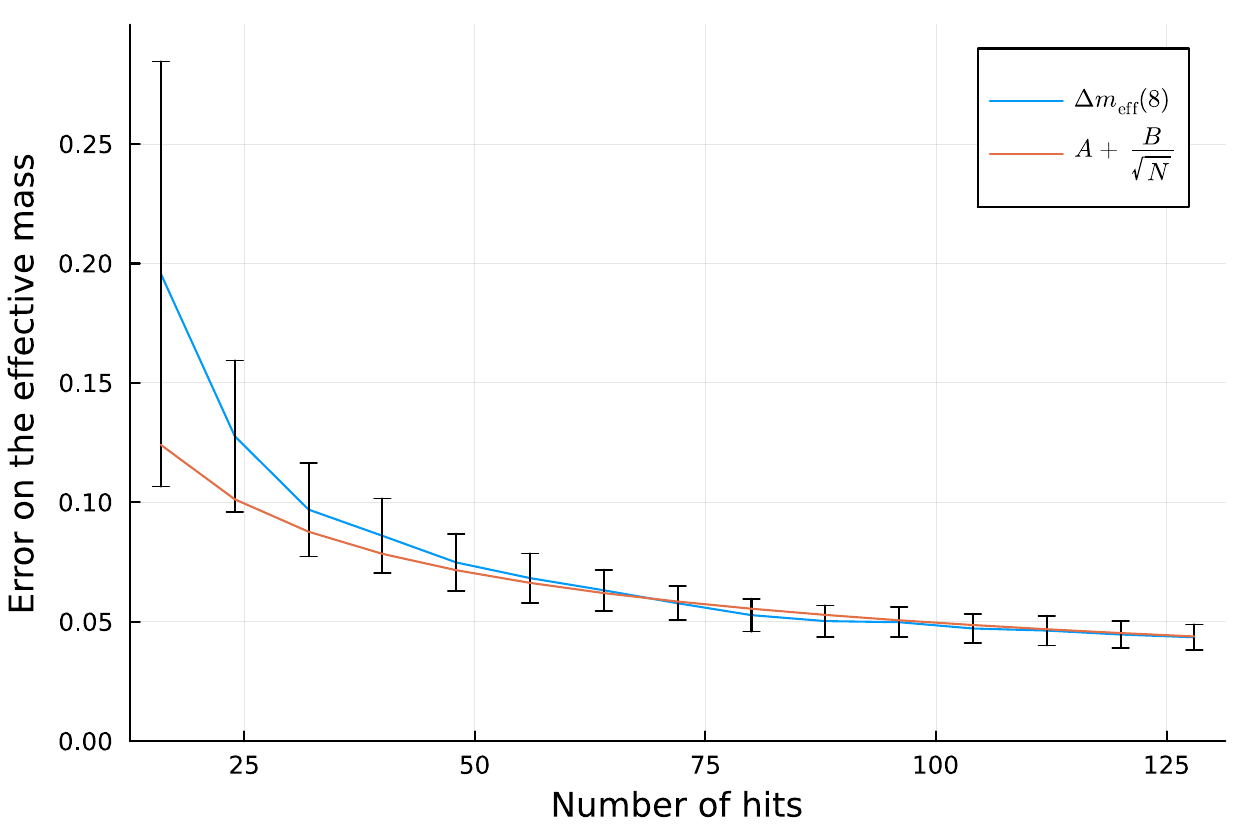}
\caption{The error on the $\sigma$ effective mass at a certain timeslice ($t = 8$) against the number of hits (stochastic inversions). The data are fitted with the simple model $\Delta m_{\mathrm{eff}}(t) = A + \frac{B}{\sqrt{\mathrm{hits}}}$. In the fit, $A = 0$, so the error in the effective mass does not appear to have plateaued with respect to the number of hits, meaning the signal is not limited by the gauge noise.}
\label{fig:stoch_err}
\end{figure}

In order to make optimum use of the gauge configurations, the number of stochastic sources used should be chosen so that the error on the effective mass is limited by the gauge noise and not the stochastic noise. To investigate which source of error is limiting the total error on the effective mass, in Figure \ref{fig:stoch_err} we plot the error on the effective mass at a single timeslice ($t = 8$) against the number of stochastic sources used, and fit a simple model. The fit suggests that we are still dominated by stochastic noise, and therefore making best use of the gauge configurations.

\newpage
\section{Conclusion}
We have described our setup for generating a chiral ensemble at $\beta = 2.2$, including the non-perturbative tuning of the action. The spectroscopy for a few key observables was reported, including the scale-setting. We then turned to the scalar singlet state, the mass of which we calculate for the first time in an ensemble generated using the exponential clover action. The effective mass is shown to agree with the mass of twice the lightest pseudoscalar state for large~$t$. This is to be expected, as the value of $\frac{\mv}{\mps}$ is around the same as in the previous setup ($2.5$)~\cite{Drach_2022}, where the $\sigma$ was shown to be stable. We are now in a good position to explore lighter mass ensembles due to the new improved action and with the new version of the HiRep suite ported to GPUs \cite{martins2024gpuacceleratedhigherrepresentationswilson, martins2024scalingsu21000gpus}. Since the ensemble in this work was the lightest that we could run on $V = 64\times 32^3$, larger volume simulations will be required. We are currently generating new, larger volume ensembles at $\beta = 2.15$ and $\beta = 2.2$ where we expect to be significantly more chiral than the ensemble described here.

\acknowledgments
This work used the DiRAC Data Intensive service (DIaL2 / DIaL3) at the University of Leicester, managed by the University of Leicester Research Computing Service on behalf of the STFC DiRAC HPC Facility (www.dirac.ac.uk). The DiRAC service at Leicester was funded by BEIS, UKRI and STFC capital funding and STFC operations grants. DiRAC is part of the UKRI Digital Research Infrastructure.

This work was also carried out using the computational facilities of the High Performance Computing Centre, University of Plymouth.

FRL is supported in part by Simons Foundation grant 994314 (Simons Collaboration on Confinement and QCD Strings), and the Mauricio and Carlota Botton Fellowship.

This project has received funding from the European Union’s Horizon 2020 research and innovation program under the Marie Skłodowska-Curie grant agreement №813942.
\bibliographystyle{JHEP}
\bibliography{bib}

\providecommand{\href}[2]{#2}\begingroup\raggedright\begin{thebibliography}{10}

\bibitem{Drach_2022}
V.~Drach, P.~Fritzsch, A.~Rago and F.~Romero-L{\'{o}}pez, \emph{{Singlet
  channel scattering in a composite Higgs model on the lattice}},
  \href{https://doi.org/10.1140/epjc/s10052-021-09914-y}{\emph{Eur. Phys. J. C}
  {\bfseries 82} (2022) 47} [\href{https://arxiv.org/abs/2107.09974}{{\ttfamily
  2107.09974}}].

\bibitem{Francis:2019muy}
A.~Francis, P.~Fritzsch, M.~L\"uscher and A.~Rago, \emph{{Master-field
  simulations of O($a$)-improved lattice QCD: Algorithms, stability and
  exactness}}, \href{https://doi.org/10.1016/j.cpc.2020.107355}{\emph{Comput.
  Phys. Commun.} {\bfseries 255} (2020) 107355}
  [\href{https://arxiv.org/abs/1911.04533}{{\ttfamily 1911.04533}}].

\bibitem{Del_Debbio_2010}
L.~Del~Debbio, A.~Patella and C.~Pica, \emph{{Higher representations on the
  lattice: Numerical simulations, $SU(2)$ with adjoint fermions}},
  \href{https://doi.org/10.1103/physrevd.81.094503}{\emph{Phys. Rev. D}
  {\bfseries 81} (2010) 094503}
  [\href{https://arxiv.org/abs/0805.2058}{{\ttfamily 0805.2058}}].

\bibitem{w0paper}
S.~Bors{\'a}nyi, S.~D{\"u}rr, Z.~Fodor, C.~Hoelbling, S.D.~Katz, S.~Krieg
  et~al., \emph{{High-precision scale setting in lattice QCD}},
  \href{https://doi.org/10.1007/JHEP09(2012)010}{\emph{JHEP} {\bfseries 08}
  (2012) 053} [\href{https://arxiv.org/abs/1203.4469}{{\ttfamily 1203.4469}}].

\bibitem{L_scher_1997}
M.~Lüscher, S.~Sint, R.~Sommer, P.~Weisz and U.~Wolff, \emph{{Non-perturbative
  $O(a)$ improvement of lattice QCD}},
  \href{https://doi.org/10.1016/s0550-3213(97)00080-1}{\emph{Nucl. Phys. B}
  {\bfseries 491} (1997) 323}
  [\href{https://arxiv.org/abs/hep-lat/9609035}{{\ttfamily hep-lat/9609035}}].

\bibitem{Sofie}
L.S.~Bowes, V.~Drach, P.~Fritzsch, S.~Martins, A.~Rago and F.~Romero-Lopez,
  \emph{{Determination of the pseudoscalar decay constant from SU(2) with two
  fundamental flavors}}, {\emph{PoS} {\bfseries LATTICE2024} (2025) 356}
  [\href{https://arxiv.org/abs/2412.06471}{{\ttfamily 2412.06471}}].

\bibitem{bowes20232flavoursu2gaugetheory}
L.S.~Bowes, V.~Drach, P.~Fritzsch, A.~Rago and F.~Romero-Lopez,
  \emph{{2-flavour SU(2) gauge theory with exponential clover Wilson
  fermions}}, \href{https://doi.org/10.22323/1.453.0094}{\emph{PoS} {\bfseries
  LATTICE2023} (2024) } [\href{https://arxiv.org/abs/2401.00589}{{\ttfamily
  2401.00589}}].

\bibitem{collaboration_2008}
P.~Boyle, C.~Kelly, R.~Kenway and A.~Jüttner, \emph{Use of stochastic sources
  for the lattice determination of light quark physics},
  \href{https://doi.org/10.1088/1126-6708/2008/08/086}{\emph{JHEP} {\bfseries
  08} (2008) 086} [\href{https://arxiv.org/abs/0804.1501}{{\ttfamily
  0804.1501}}].

\bibitem{martins2024gpuacceleratedhigherrepresentationswilson}
S.~Martins, E.~Kjellgren, E.~Molinaro, C.~Pica and A.~Rago,
  \emph{{GPU-accelerated Higher Representations of Wilson Fermions with
  HiRep}}, \href{https://doi.org/10.22323/1.451.0035}{\emph{PoS} {\bfseries
  EuroPLEx2023} (2024) 035} [\href{https://arxiv.org/abs/2405.19294}{{\ttfamily
  2405.19294}}].

\bibitem{martins2024scalingsu21000gpus}
S.~Martins, E.~Kjellgren, E.~Molinaro, C.~Pica and A.~Rago, \emph{{Scaling
  SU(2) to 1000 GPUs using HiRep}}, {\emph{PoS} {\bfseries LATTICE2024} (2025)
  } [\href{https://arxiv.org/abs/2411.18511}{{\ttfamily 2411.18511}}].

\end{thebibliography}\endgroup

\end{document}